\documentclass[namedreferences]{solarphysics}
\usepackage[optionalrh,solaromanenum]{spr-sola-addons} 
\usepackage{graphicx}                    
\usepackage{color}                       
\usepackage{url}                         

\newcommand{\apj}{    {Astrophys. J.}\ }
\newcommand{\apjl}{   {Astrophys. J. Lett.}\ }

\newcommand{\mnras}{  {Mon. Not. Roy. Astron. Soc.}\ }

\newcommand{\pasj}{   {Pub. Astron. Soc. Japan}\ }

\newcommand{\solphys}{{Solar Phys.}\ }

\newcommand{\memsai}{    {Mem. Soc. Astron. Italiana}\ }

\begin{document}

\begin{article}

\begin{opening}


\title{Transient artifacts in a flare observed by the Helioseismic and Magnetic Imager on the Solar Dynamics Observatory}


\author{J.C.~\surname{Mart\'inez Oliveros}$^{1}$\sep
C.~\surname{Lindsey}$^{2}$\sep 
H.S.~\surname{Hudson}$^{1,3}$\sep
J.C.~\surname{Buitrago Casas}$^{4}$
}

\runningtitle{Transient artifacts in a flare observed by SDO/HMI}
\runningauthor{J.C. Mart\'{\i}nez Oliveros \textit{et al}.}


\institute{
$^{1}$ Space Sciences Laboratory, University of California, Berkeley, CA USA;  email: \url{oliveros@ssl.berkeley.edu}, \url{krucker@ssl.berkeley.edu}, \url{hhudson@ssl.berkeley.edu}\\
$^{2}$ North West Research Associates, CORA Division, Boulder, CO USA;
email: \url{clindsey@cora.nwra.com}\\
$^{3}$ Department of Physics \& Astronomy, University of Glasgow, Glasgow, Scotland, UK\\
$^{4}$ Observatorio Astron\'ominco Nacional, Universidad Nacional de Colombia, Bogot\'a, Colombia;
email: \url{jcbuitragoc@unal.edu.co}
}


\begin{abstract}
The \textit{Helioseismic and Magnetic Imager} (HMI) onboard the \textit{Solar Dynamics Observatory} (SDO) provides a new tool for the systematic observation of white-light flares, including Doppler and magnetic information as well as continuum. In our initial analysis of the highly impulsive $\mathrm \gamma$-ray flare SOL2010-06-12T00:57 (\citeauthor{2011SoPh..269..269M}, \textit{\solphys}, \textbf{269}, 269, \citeyear{2011SoPh..269..269M}), we reported the signature of a strong blueshift in the two footpoint sources.  Concerned that this might be an artifact due to aliasing peculiar to the HMI instrument, we undertook a comparative analysis of \textit{Global Oscillations Network Group} (GONG++) observations of the same flare, using the \textit{PArametric Smearing Correction ALgorithm} (PASCAL) algorithm to correct for artifacts caused by variations in atmospheric smearing.  This analysis confirms the artifactual nature of the apparent blueshift in the HMI observations, finding weak redshifts at the footpoints instead.  We describe the use of PASCAL with GONG++ observations as a complement to the SDO observations and discuss constraints imposed by the use of HMI far from its design conditions. With proper precautions, these data provide rich information on flares and transients.
\end{abstract}



\end{opening}



\section{Introduction}

Solar flares are well-known phenomena characterized by what we understand to be a sudden release of energy stored in coronal magnetic fields. It is widely accepted that solar flares are driven by magnetic reconfiguration, suggesting that the evolution of the magnetic field in the interior of the Sun may have clear and visible consequences in the solar atmosphere. These effects can be seen at different heights, revealing important physical processes related to the dynamics of the flare itself. In the chromosphere and photosphere, flares are seen as sudden brightness increases in the hydrogen absorption lines.  In some instances there is excess visible-continua emission, in what is called the ``white-light flare" phenomenon. 

The first solar flare was discovered by \inlinecite{1859MNRAs..20...13C} by its excess emission of white light. The intensity of excess white light in some flares represents the release of considerable energy in the lower atmosphere in a few seconds or minutes. In the upper atmosphere and corona, X-ray signatures indicate the existence of electrons accelerated to tens of keV.  In some flares, $\gamma$-rays may also appear, and these reveal the presence of protons with energies in excess of some 10~MeV. In white-light flares, the onset of the continuum emission is in the impulsive phase.  Its source is deep in the solar atmosphere, extending into the near-IR spectrum, including even  the ``opacity minimum'' at $\approx$\,1.5~$\mu$m \cite{2004ApJ...607L.131X}. In spite of the photospheric origin of most of this part of the spectrum in non-flaring conditions, strong evidence implicates the chromosphere for at least part of the excess in white-light flares.  The continuum emission has signatures of recombination radiation \cite{1986lasf.conf..142N,2010MmSAI..81..637H}.  Moreover, the source of the visible continuum appears to correlate well with hard X-rays \cite{1970SoPh...13..471S,1975SoPh...40..141R,1992PASJ...44L..77H,2012ApJ...753L..26M}.  The \textit{Helioseismic and Magnetic Imager} (HMI) onboard the \textit{Solar Dynamics Observatory} (SDO) gives us a new opportunity to study the white-light enhancement produced by flares and explore the close relation it has with the energy deposition observed in hard X- and $\gamma$-rays. This is possible because of the spatial and spectral capabilities of HMI \cite{2012SoPh..275..229S}.  The HMI data clearly resolve the profile of the Fe~{\sc i}~line at 6173.34~\AA , in each $\approx$\,0.5$''$ pixel and 45-second time step.

\begin{figure}[htbp]
\begin{center}
  \includegraphics[width=1.\textwidth]{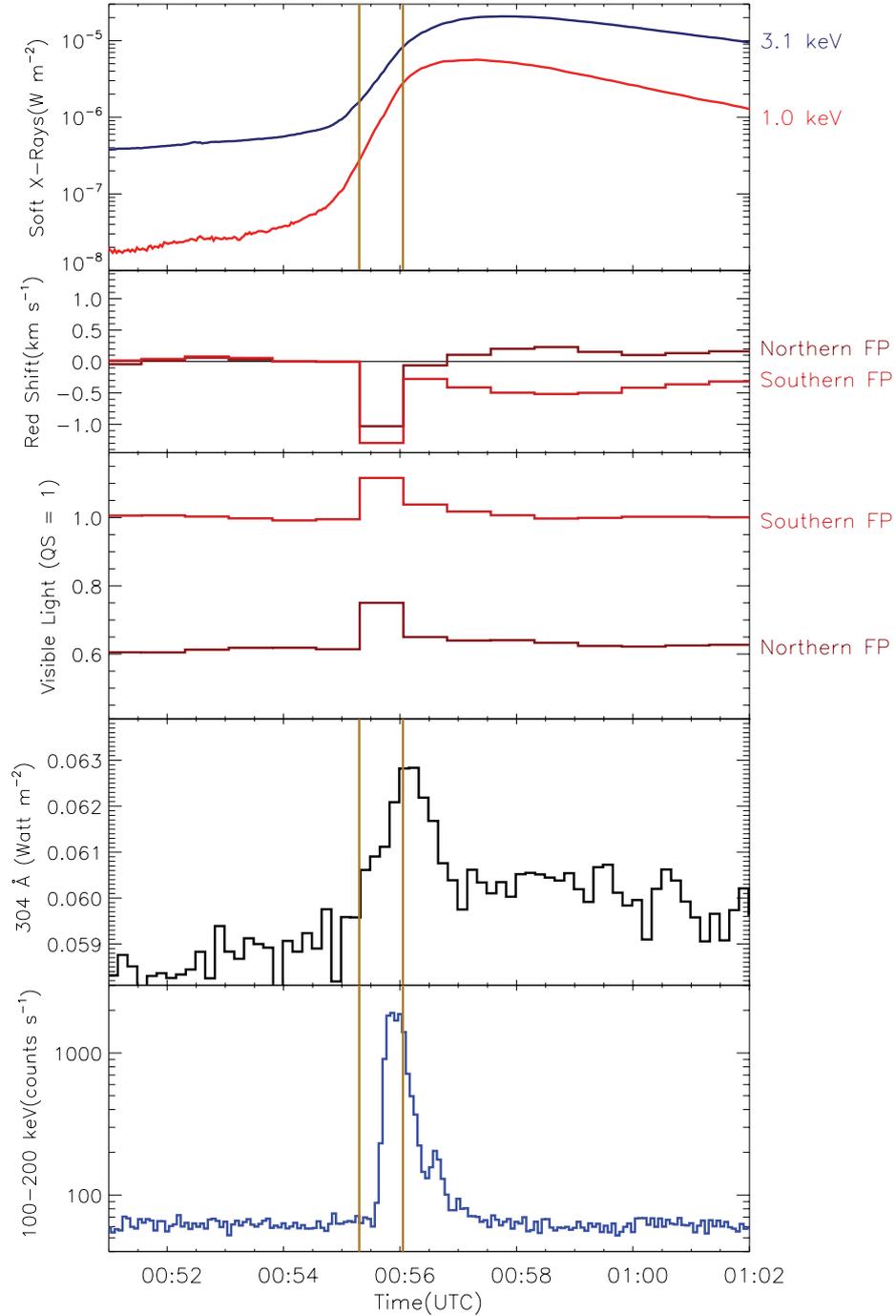}
\label{fg:overview}
\caption{Overview of the temporal development of SOL2010-06-12T00:57. The five panels show (top to bottom): GOES soft X-rays, HMI NRT (``near real time") redshifts in the two flaring footpoints, continuum brightnesses in the two regions, EVE 304\,\AA\, excess irradiance, and RHESSI \protect\cite{2002SoPh..210....3L} 100\,--\,200 keV hard X-ray flux. The vertical lines show the beginning and end of the sampling interval of the HMI observations used to construct the Doppler signature (see also Table 1). Reproduced by permission from \protect\inlinecite{2011SoPh..269..269M}.}
\end{center}
\end{figure}

The present study recognizes such an instance in a study by \inlinecite{2011SoPh..269..269M}, herein Paper I, of the highly impulsive M2.2-class flare of SOL2010-06-12T00:57. During the impulsive phase of this flare, a bright white-light kernel appeared in each of the two magnetic footpoints revealed by hard X-ray emission. When the flare occurred, the spectral coverage of the HMI filtergrams (six equidistant samples spanning $\pm$172~m\AA\ around nominal line center) encompassed the line core and the blue continuum sufficiently far from the core to eliminate significant Doppler crosstalk in the latter, which is otherwise a possibility for the extreme conditions in a white-light flare. The Fe~{\sc i}~line appears to be shifted to the blue during the flare but does not go into emission; the contrast is nearly constant across the line profile.

In this study, then, we show that Doppler signatures in SDO/HMI observations of white-light flares, while excellent in many other respects, are subject to strong artifacts, because different parts of the spectrum are recorded at different times, between which the respective intensities undergo significant variations.  The instruments used by the \textit{Global Oscillations Network Group} (GONG) are essentially free of these artifacts, because all parts of the spectrum are effectively measured at the same time. Even so, this does not imply that the GONG++\footnote{GONG++ is an upgrade of the GONG 256x256 to 1024x1024 pixel cameras (known as GONG+, \opencite{1998ESASP.418..209H}) and its corresponding data processing system.} observations are correct or impervious to other kinds of artifacts, as demonstrated by \inlinecite{2010SoPh..262..337M}. We should mention that artifacts in HMI data have been reported before (\textit{e.g.} \opencite{2012ApJ...747..134M}), but, the nature of these artifacts we believe to be different, and to have similar properties to those reported by \inlinecite{2002ApJ...565.1335Q} and \inlinecite{2003ApJ...599..615Q}. 

The major disadvantage of the GONG++ observations is the smearing of the images due to atmospheric turbulence.  This limits spatial resolution.  However, more troubling, variations in this smearing from one image to the next introduces strong artifacts in features that have strong spatial intensity gradients, such as active regions.  We have developed a ``PArametric Smearing Correction ALgorithm" (PASCAL) to correct these artifacts \cite{2008SoPh..251..627L} and applied it to GONG++ observations of SOL2010-06-12T00:57 for comparison with the SDO/HMI Doppler signatures.  The GONG++ Doppler observations, while highly consistent with the HMI Doppler observations in non-flaring conditions, show no facsimile of the strong blueshift indicated by the HMI Doppler observations in the impulsive-phase foot points of the flare.  On the contrary, they rather show a weak redshift at those locations at that time (see Section~\ref{sec:observations}).  Figure~\ref{fg:overview} compares the HMI Doppler and intensity signatures temporally with UV and X-ray observations from \textit{Geostationary Operational Environmental Satellite} (GOES), \textit{Reuven Ramaty High Energy Solar Spectroscopic Imager} (RHESSI) and \textit{Extreme ultraviolet Variability Experiment} (EVE).

\section{The PArametric Smearing Correction ALgorithm (PASCAL)}
\label{sec:gong}

The GONG++ network provides us high-quality Doppler and white-light observations of the Sun. The significant limitation of these data for active-region observations is the smearing of the solar image due to atmospheric turbulence. This imposes familiar limits on spatial resolution.   For active-region seismology, the main liability of image smearing is noise in the helioseismic signature due to variations in the amount of smearing from one image to the next. In the quiet Sun, these variations are small compared to the signatures that GONG++ is designed to image. However, the effect of temporal variations in the smearing acting on the large intensity gradients intrinsic to sunspots, for example, introduce noise that competes with helioseismic signatures.
Until recently, this has made seismology of active regions from  GONG++ data impractical except under unusually good atmospheric conditions \cite{1993ApJ...415..847T}.

\inlinecite{2008SoPh..251..627L} devised an algorithm that discriminates most of this noise from helioseismic signals, making active-region seismology based on GONG++ observations practical on terms similar to that for the quiet Sun in moderately good atmospheric conditions.  The method is specifically applicable to observations that are integrated over a time that is long, one minute for the GONG++ observations, compared to the characteristic time for atmospheric turbulence in the telescope beam, typically $\approx$30~ms, with high tracking accuracy based on location of the solar limb. Under these conditions, the resulting smearing is highly isotropic in the image plane. The noise that competes with seismic observations in an active region is the result of differential variations in the degree of smearing from one integration to the next. The method that \inlinecite{2008SoPh..251..627L} developed applies a measured ``adjustment'' in smearing to each GONG++ image to devise a resultant smear that is effectively the same for all images. The method is greatly simplified by the recognition that the difference between two images can be well represented by a constant times the Laplacian of the average of the two images. \inlinecite{2008SoPh..251..627L} describe the method analytically, applying it successfully to GONG++ intensity observations of the white-light flare SOL2003-10-29.

\inlinecite{2008SoPh..251..627L} successfully applied the technique to GONG++ continuum images. More recently its extension to Doppler helioseismic applications has been demonstrated by \inlinecite{2011ApJ...739...70Z}. Because GONG++ Doppler and magnetic images are linear combinations of intensity maps in different wavelengths and polarizations, and the Laplacian is a linear operator, the extension of the technique to  GONG++ Doppler and magnetic images is straightforward.

For a given time series of $n$ GONG++ intensity images, indexed by $i \in \{1, 2, 3, ..., n\}$, this entails first deriving a set $\sigma_i$ of smearing parameters that characterizes each respective intensity image. The smearing of each intensity, Doppler, and magnetic image is then adjusted differentially to eliminate stochastic variations in $\sigma_i$ from one to the next. It is important to understand that the foregoing procedure will not substantially recover high spatial resolution lost as a result of atmospheric turbulence. It only eliminates spurious intensity variations due to differential variations in the amount of smearing. This procedure will make it possible to apply the same helioseismic techniques to active regions in GONG++ observations as in the concurrent space-borne \textit{Michelson Doppler Imager} (MDI) observations. This includes the large inventory of acoustically active flares that MDI onboard the \textit{Solar and Heliospheric Observatory} (SOHO) will have to have missed because of limited ($\sim$20\,\%) temporal coverage of Cycle 23 after the advent of GONG++, as well as the SDO/HMI events.

\begin{figure}[htbp]
\begin{center}
  \includegraphics[width=1\textwidth]{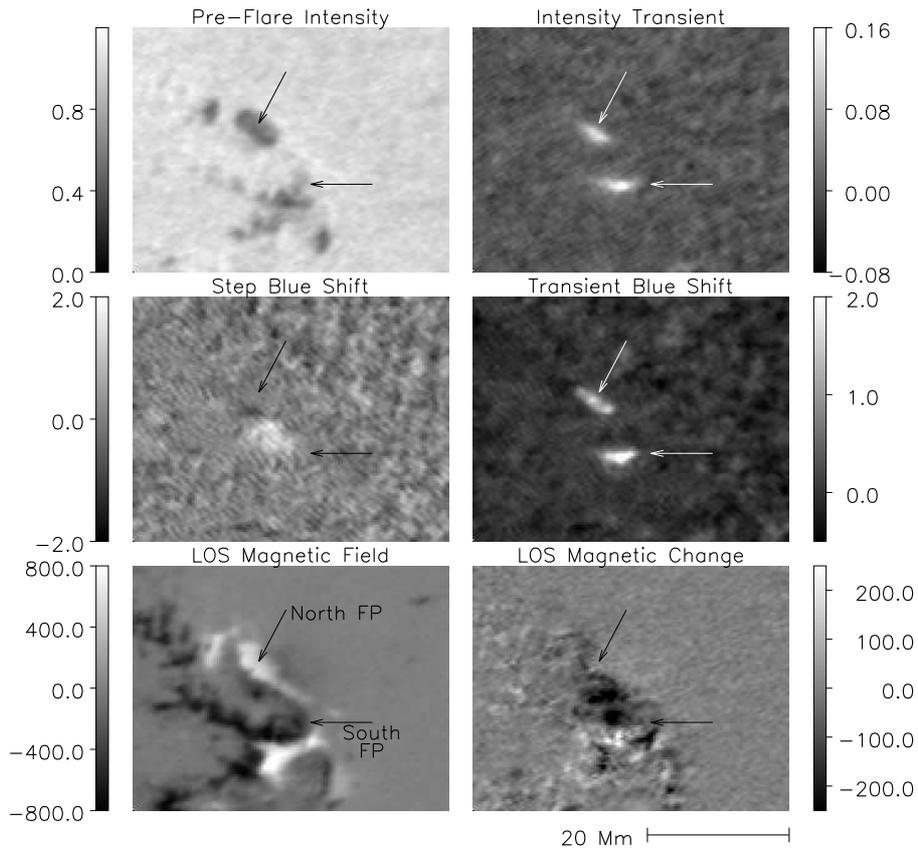}
\label{fg:maps}
\caption{HMI ``near real time" (NRT) images: top-left, HMI preflare continuum image; top-right, transient continuum excess over the preflare continuum; middle-left, Doppler difference between pre- and post-flare intervals, showing a stepwise change in line-of-sight flow; middle right, map of blueshift in the impulsive phase minus the same 45 s earlier; lower left, the preflare line-of-sight magnetic field; and lower-right, the Postflare line-of-sight magnetic field minus the preflare line-of-sight field (see Table~1 for times). The arrows locate the footpoint sources. The grayscale shows fractions of quiet-Sun intensity, [$\mathrm{km\,s^{-1}}$], and Gauss for the top, middle and bottom rows, respectively. }
\end{center}
\end{figure}

\section{Observations and Analysis}
\label{sec:observations}

In Paper~I we studied the M2.0-class flare SOL2010-06-12T00:57 (NOAA 11081, located approximately at N22W45). This flare had a remarkably impulsive hard X-ray light curve, with a duration (half maximum at 50~keV) of only about 25~seconds. It also produced $\gamma$-ray emission, unusual for such a weak event \cite{2009ApJ...698L.152S}. Figure~\ref{fg:overview} compares time series representing HMI intensity and Doppler observations with soft (GOES) and hard (RHESSI) X-ray fluxes.  
Table~1 summarizes these phenomena seen in the relevant HMI intervals and gives them names, ``Preflare,'' ``Brightening,'' and ``Postflare,'' for subsequent reference. The hard X-ray signature mainly matches the Brightening interval (the white-light flare) and does not extend earlier into the previous HMI integration.  

\begin{table*}[h]
\caption{Flare timeline in HMI data (45-second data frames)}
\centering
\smallskip
\begin{tabular}{l l l l l}
Interval name & Start [UTC] & HMI continuum & HMI Doppler & RHESSI  100~keV\\
\hline
Preflare & 00:54:11 & no excess & no anomaly & none \\
Brightening & 00:55:41 & $>$10\,\% increase & 2~km s$^{-1}$ blue & bright \\
Postflare & 01:00:11 & no excess & complex & no detection \\
\hline
\end{tabular}
\end{table*}
\label{tab:signs}

\begin{figure}[htbp]
\begin{center}
\hspace{-1.cm}
  \includegraphics[width=1.075\textwidth]{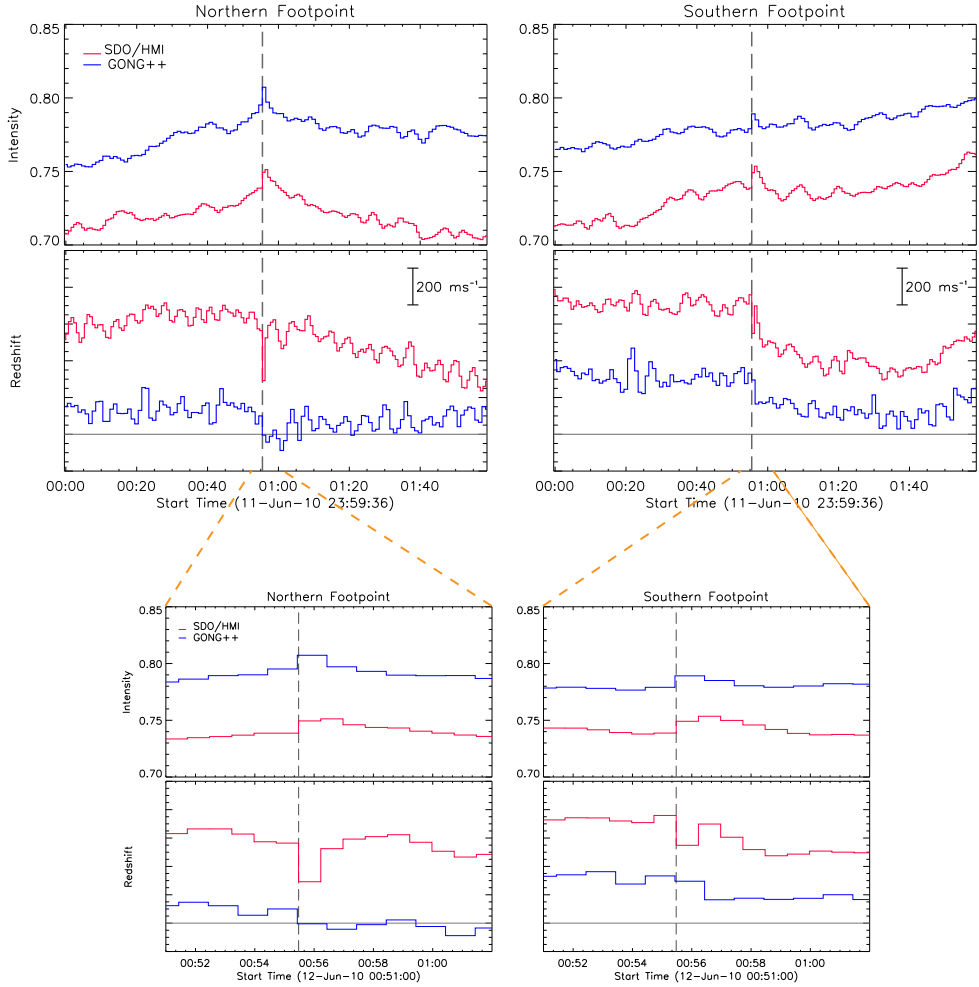}
\label{fg:pascal}
\caption{Comparison of GONG++ and HMI observations of SOL2010-06-12, isolating the two flare footpoints as resolvable by GONG++.  Upper: intensity contrast relative to a quiet-Sun reference; lower: redshift. SDO/HMI data are in red, GONG++ in blue. The striking blue-shifted Doppler artifact seen by HMI in the impulsive phase (dashed vertical line) is absent in the GONG++ Doppler signatures.  Note the strong  correlations between the GONG++ and HMI time series in the intensity and Doppler time series other than in the Doppler time series during the impulsive phase. The bottom four frames show a closeup at the time of the artifact.}
\end{center}
\end{figure}

Figure~\ref{fg:maps} shows maps of continuum, Doppler and line-of-sight magnetic signatures of the flare.  The top-left frame shows the Preflare continuum intensity; the top-right frame shows the continuum excess during the Brightening phase; the middle-left frame shows the Doppler variation from Preflare to Postflare. The middle-right frame shows the same for Preflare to Brightening, the bottom left frame shows the Preflare line-of-sight magnetic field; the bottom-right frame shows its variation from Preflare to Postflare. Paper~I noted strong Doppler-blueshift transients in both footpoints (middle-right frame of Figure~\ref{fg:maps}).  Timeseries of these signatures are plotted in Figure~\ref{fg:pascal}. In Paper~I the validity of this blueshift in the HMI Dopplergrams was questioned, but left the question unresolved.  Such a blueshift, if real, could open significant issues to flare physics.  Among the possible implications are rapidly up-flowing photospheric gas with the line in absorption; rapidly down-flowing chromospheric gas with the line from the chromospheric contribution in emission (but the composite line still in absorption); and rapid horizontal flows, noting that the line of sight at the foot points was inclined $\approx$50$^\circ$ from vertical.  The significant concern is that the intensity maps sampling different wavelengths are observed at different times.  If the respective intensities change considerably in the respective interval, this will introduce a spurious Doppler signature even if there is never any spectral shift at any single moment during the sampling interval.

The most practical control for the validity of the Doppler transient seems to be by comparing the results of the analysis with other instruments. \inlinecite{2011SoPh..273...69H} show that for this event the EVE observations report a redshift of $16.8 \pm 5.9$~$\mathrm{km\,s^{-1}}$ in the He~{\sc ii}~304-\AA\ line. This might possibly suggest that the blue Doppler transient is an artifact, but the gas represented by this line is in the transition region, the line being in emission rather than absorption.  Indeed, a layer of down-flowing gas in the chromosphere or upper photosphere that by itself would appear in emission would manifest a {\it blue} shift if the composite line remained in absorption.  So, this is not a practical control for the photospheric Doppler signatures.

What is needed is a comparison of the HMI observations with observations from an instrument that observes a photospheric line similar to that observed by HMI, but that samples all parts of the spectrum simultaneously.  The GONG++ network provides such an instrument.  Each instrument in the network is a  Michelson Doppler tachometer that measures the spectral shift of the photospheric line Ni~{\sc i}~6768~\AA, with an integration time of 60 seconds \cite{brown1984,1996Sci...272.1284H,1995ASPC...76..381L,1996Sci...272.1284H}.  SDO/HMI observes the slightly weaker line, Fe~{\sc i}~6173.34~\AA\ with a 45-second cadence.  This line features greater magnetic sensitivity, with the caution \cite{norton2006} that this reduces the dynamic range over which HMI can reliably measure Zeeman splitting, given the limited spectral range sampled by the instrument and the considerable part of this range covered by solar rotation and motion of the spacecraft even without magnetic splitting.

The heights at which these lines form are significantly different, but the contribution functions overlap strongly. \inlinecite{2011SoPh..271...27F} studied the formation heights of the Fe~{\sc i}~6173-\AA\ (HMI) and Ni~{\sc i}~6768-\AA\ (MDI, GONG++) lines based on 3D radiation-hydrodynamic simulations. They found that the Fe~{\sc i}~6173-\AA\ line is formed at a height $\approx$100~km, while the Ni~{\sc i}~6768-\AA\ line is formed at $\approx$150~km. Also, they studied the dependency of the height of formation on the spatial resolution, finding a small dependence that increases the height by about 40\,--\,50~km for HMI and 15\,--\,25~km for MDI. As a result the formation height of HMI Fe~{\sc i}~6173-\AA\ and MDI Ni~{\sc i}~6768-\AA\ lines become 140-150~km and 165\,--\,175~km, respectively. (For more details see \opencite{2011SoPh..271...27F} and references therein).

A significant consideration in a comparison between GONG++ and HMI is the disparity in spatial resolution. SDO/HMI has far better spatial resolution than GONG++.  In order to compare the footpoints in each instrument, the HMI images were rebinned to the GONG++ resolution, so that we could do the same analysis for both data sets. We applied the PASCAL algorithm to the GONG++ observations to correct for variable smearing by the Earth's atmosphere.  We then calculated the intensities and Doppler signatures averaged spatially over masks that encompassed the respective foot points in both GONG++ and HMI images.  Because of the poorer spatial resolution in the GONG++ observations, the masks applied in this analysis were more extended than those applied in Paper~I. The assumption at this moment is that if the HMI blueshift is valid, it should be clearly seen in the GONG++ Doppler time series. Figure~\ref{fg:pascal} shows the intensity and velocity time series for both instruments. Both instruments show similar behaviors in the intensity data.  Moreover, there is a strong correlation between the respective Doppler time series before and after the impulsive phase of the flare.  The strong blueshifts reported in Paper~I in the impulsive phase of the flare (vertical dashed line), although weaker when averaged over the more extended masks, remain clear in both of HMI Doppler time series.  However, these features do not appear in the GONG++ Doppler time series.  This points to the presence of a strong artifact in the HMI Dopplergrams when applied to transient phenomena such as those that occur in flares. We note again, that these results do not demonstrate the reliability of GONG++ data during flares, and only shows that the observed Doppler transient is an artifact  in the HMI Dopplergrams.

\section{Conclusions}

We find that SDO/HMI Doppler observations, while excellent for nominal helioseismic applications, are subject to strong artifacts when applied to transient phenomena, such as occur in flares, as anticipated.  This appears to be the result of rapid changes of the spectrum between the respective times at which different parts of it are being sampled.  Because the GONG++ instrument effectively samples all parts of the spectrum simultaneously, it appears to be immune to Doppler artifacts of this kind, notwithstanding that both GONG++ and HMI remain subject to effects due to various aspects of radiative transfer in the flare environment and independent of considerations of relative timing.  Similar considerations ought to apply to magnetograms, both line-of-sight and Stokes, during the impulsive phases of flares as determined by  \inlinecite{2010SoPh..262..337M}.  We know of no reason that these artifacts should be of any significant concern for non-flaring applications of GONG++ or HMI observations, such as standard $p$-mode analysis (\textit{e.g.} \opencite{basu2003}; \opencite{rajaguru2004}; \opencite{barban2004}; \opencite{hughes2005}; \opencite{moradi2006}; \opencite{2011ApJ...739...70Z}).  

That GONG++ has been working for a little over a decade and half, and it has been possible to improve and refine its methods for obtaining and analyzing helioseismic observations, reinforces its reliability for the range of applications to which it has been applied, which now include white-light flares along with the nominal helioseismic applications.  The major disadvantage of GONG++ as a complement to SDO/HMI for transient phenomena is its poorer spatial resolution.  Drifts in atmospheric smearing have made active-region seismology impractical until recently, with the development of the PASCAL algorithm.  The application of PASCAL to GONG++ intensity observations has made it useful for the recognition and analysis of white-light flares (\opencite{2008SoPh..251..627L}; Paper~I; this study).  More recent applications to GONG++ Doppler observations by \inlinecite{2011ApJ...739...70Z} have now estabished its utility for active-region seismology.  These results argue in favor of a strong role for GONG++ as a complement to SDO/HMI in the rapidly developing field of flare seismology.

\bigskip\noindent{\bf Acknowledgments:}
We thank Jesper~Schou and Sebastien~Couvidat for their valuable comments during the writing of this article. The Berkeley group was supported by NASA under contract NNX11AP05G, and the SDO/HMI by contract NAS5-02139 to Stanford University. These data have been used courtesy of NASA/SDO and the EVE and HMI science teams. This work utilizes data obtained by the Global Oscillation Network Group (GONG) Program, managed by the National Solar Observatory, which is operated by AURA, Inc. under a cooperative agreement with the National Science Foundation.


\end{article} 
\end{document}